\documentclass[12pt]{iopart}

\usepackage{iopams}
\usepackage{amsfonts}   
\usepackage{amssymb}   
\usepackage{float}
\usepackage{subfig}
\usepackage{graphicx}

\begin{document}

\title[planar micro Penning trap]{Fabrication of a planar micro Penning trap and
numerical investigations of versatile ion positioning protocols}

\author{M. Hellwig$^{(a)}$, A. Bautista-Salvador$^{(a)}$, K. Singer$^{(a)}$,
G. Werth$^{(b)}$, F. Schmidt-Kaler$^{(a)}$}

\address{(a) Institut f\"ur Quanteninformationsverarbeitung, Universit\"at Ulm, Albert-Einstein-Allee 11,
D-89069 Ulm, Germany, \\
(b) Physikalisches Institut, Johannes-Gutenberg-Universit\"at, 55099 Mainz, Germany}

\ead{ferdinand.schmidt-kaler@uni-ulm.de}

\begin{abstract}

We describe a versatile planar Penning trap structure, which allows to dynamically
modify the trapping configuration almost arbitrarily. The trap consists of 37
hexagonal electrodes, each with a circumcirle-diameter of 300~$\mu$m,
fabricated in a gold-on-sapphire lithographic technique. Every hexagon can be
addressed individually, thus shaping the electric potential. The fabrication of
such a device with clean room methods is demonstrated. We illustrate the
variability of the device by a detailed numerical simulation of a lateral and a
vertical transport and we simulate trapping in racetrack and artificial crystal
configurations. The trap may be used for ions or electrons, as a versatile
container for quantum optics and quantum information experiments.
\end{abstract}

\pacs{37.10.Ty, 03.67.Lx}

\maketitle

\section{Introduction}

Trapped ions have been an important workhorse for quantum information research, even
though research has been mostly limited to one dimensional arrangements of ion crystals
in radio frequency Paul traps \cite{HAFF08,BLATT08}. Planar Paul traps have shown their
advantages for scalable micro fabricated devices
\cite{MAD04,PER06,SEID06,LEIB07,WANG09,AMI08}, even for trapping cold molecules
\cite{WES08}. However a two-dimensional arrangement of ions \cite{SCHMIED09} still needs
to be shown. Inspired by experiments with large, rotating planar ion crystals in
Penning traps \cite{SKI04}, as well as single ions \cite{CRI08} or electrons
\cite{GAB86,HANN08}, novel designs for Penning traps have been proposed
\cite{CA05,CA07,STA05}. As an advantage  \cite{TAY07},
Penning traps operate with only static electric and magnetic fields, thus noise
and heating issues as observed for radio frequency ion traps \cite{LAB08} are expected
to be largely suppressed. The AC drive of Paul traps, with frequencies from 1 to 100~MHz, and with peak-to-peak voltages in the range of a few 100~V constitutes a considerable source of noise, which affects and heats the trapped particles via free charges and surface patches. The Penning trap, however, even when ions are trapped in the vicinity of slightly contaminated surfaces, still is expected to exhibit much better noise properties, as all fields including the patch fields are static.

The static magnetic field lifts the degeneracy of ground
state spin states, allowing for a natural choice of qubit basis states.
Currently, one of the most urgent unsolved scientific tasks in our field
concerns building a scalable quantum information processing device. In this
letter we show that novel planar and micro structured Penning traps allow for
trapping various qubit configurations including two-dimensional ion crystals
with well controlled interactions between individual qubits and with individual
initialization and readout. Cluster state generation and quantum simulation are
future applications of such tailored qubit arrays.

The paper first describes a pixel micro structure to generate the complex electric
potential for the novel Penning trap. We briefly sketch the numerical method to
calculate the potential landscape with the required high accuracy and optimize the
control voltages for a selected set of qubit configurations and qubit operations.  We
estimate two-qubit spin-spin interactions \cite{JOH09,DAN09}, which would lead either
to a multi-qubit cluster state \cite{WUN09} or the realization of plaquette physics
\cite{PAR08} in a rich, however well controlled quantum optical few-spin-system. In the
last section we show the fabrication of such a device with state of the art clean room
technology.

\section{Versatile trapping configurations and numerical simulations}

The genuine three dimensional Penning trap \cite{GHO1996,GAB86,HANN08} is composed of a quadrupole electrical potential $\Phi^{el.}$ provided by a voltage $U$ applied between a ring electrode and two electrically isolated end cap electrodes of hyperbolic shape,
and a superimposed constant magnetic field B$_0$ in the direction of the $z$-axis.
\begin{eqnarray}
  \Phi^{el.}(\rho,z) &=& \frac{U}{r_0^2}(\rho^2-2z^2)
\end{eqnarray}
The characteristic dimension of the trap electrodes is denoted by $r_0$. The electric field confines charged particles in the axial-direction (along the $z$-axis) while the magnetic field prevents them from escaping in the radial direction $\rho$. The motion in this potential consist of three harmonic quantum oscillations at the frequencies
\begin{eqnarray}
  \omega_{\pm} &=& (\omega_c \pm \sqrt{\omega_c^2 - 2 \omega_z^2})/2 \\
   \omega_z &=& \sqrt{2eU/md^2}
\end{eqnarray}
where $\omega_c = e B_0 / m$ is the free cyclotron frequency, 400~kHz/T in case of $^{40}\mathrm{Ca}^+$ ions, which means that in typical experimental settings with singly charged ions \cite{SKI04,CRI08} one may reach a few MHz of cyclotron frequency. In the case of trapped electrons \cite{GAB86,HANN08,BUSH08} the cyclotron frequency easily exceeds even 100~GHz. $\omega_+$ is called the reduced cyclotron frequency, $\omega_-$ the magnetron frequency, and $\omega_z$ the axial frequency. We note that in the radial direction the ion moves on a potential hill, making the magnetron motion metastable, where the energy of the magnetron motion decreases with increasing quantum number. Axialization, for example by using a rotating electric field (see sect.~\ref{transport sect}), excites the ion thereby moving it towards the center. The Penning trap stability condition  $2\omega^2_{ax} <  \omega^2_{cyc}$ (see, e.g., \cite{GHO1996}) limits the maximum axial confinement frequency.

\begin{figure}[h]
\begin{center}
\includegraphics[width=1\textwidth]{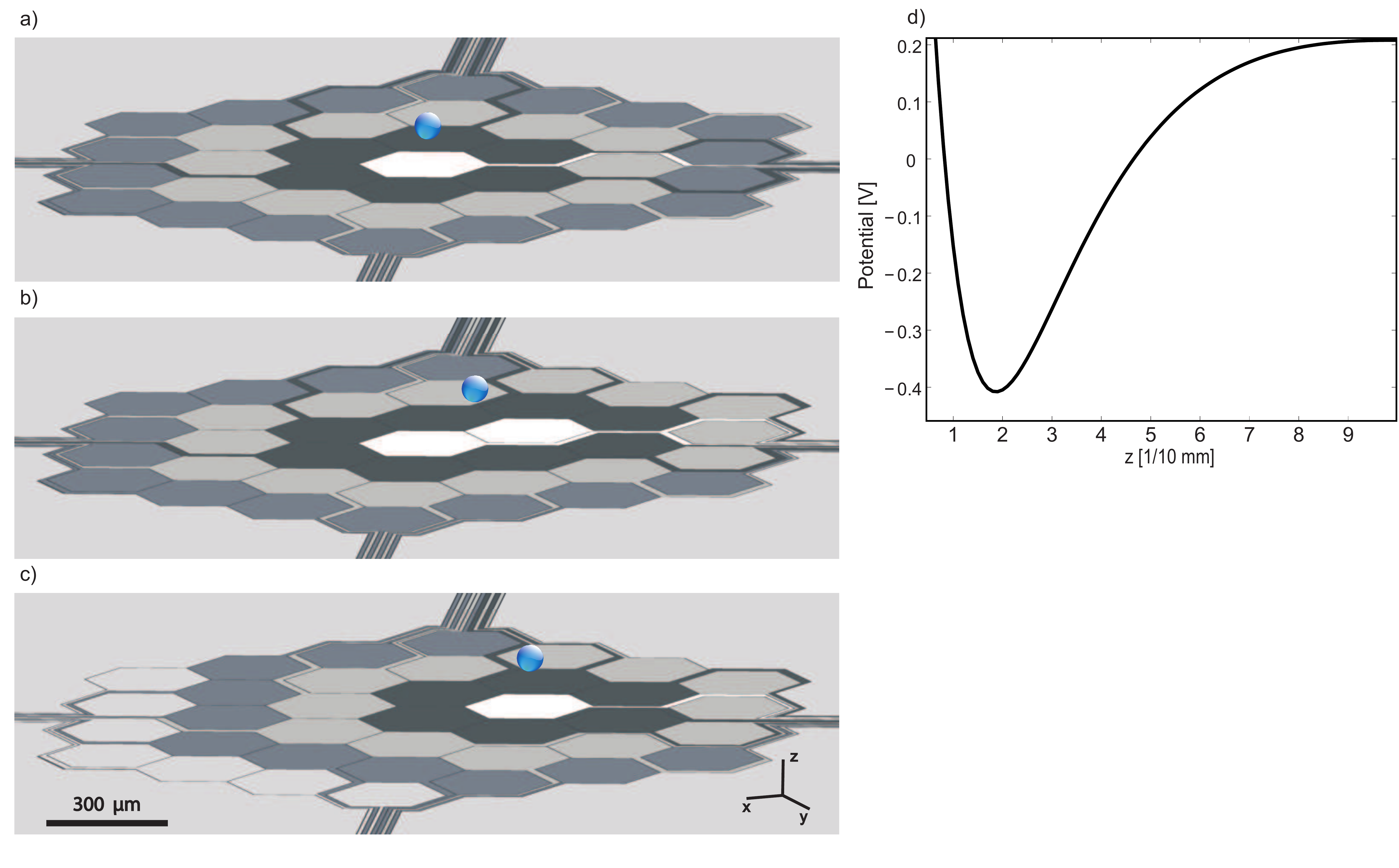}
\caption[]{a) Schematics of the Pixel trap for single ion trapping and
b,~c) for a lateral transport of an ion by changing of applied potentials. d) Potential
$\Phi_{el}(z)$ relevant for the axial confinement. If we chose addressing the pixels
with $1$~V: white, $-2.8$~V: black, $1$~V: light grey, and $3$~V: dark grey, we result in an
axial frequency of $\omega_{ax}/(2\pi)$ = $1.6$~MHz for calcium ions. The
magnetic field points out of orthogonal to the plane of electrodes.}
\label{iontrapping}
\end{center}
\end{figure}

In our case the electric field is generated by surface electrodes in one plane \cite{STA05} instead of the three dimensional ring structure, see Fig.~1. Apart from this construction feature, the motional degrees of freedom are similar to those in a three dimensional Penning trap. In a complex geometry such as our Pixel trap with a large number of different electrodes to control the confinement, see Fig.~\ref{iontrapping}a, numerical simulations are used to find a set of voltages for specific tasks and configurations with a few ion qubits. For the shape of the control electrodes, we decided to use a hexagonal shape
with a circumcircle diameter of 300$\mu$m, which allows the forming of spherically symmetric potential field geometries.

In order to obtain accurate electrostatic potentials with minimum numerical
noise for our geometry we apply the fast multipole method for the boundary
element problem and calculate the potentials with the exact
expressions\cite{SingerRMP2009}. Therefore a mesh of $N\sim 12446$ surface
rectangles is placed on the surface geometry and the surface charges are then
obtained in only $ O(N)$ steps. As opposed to finite elements or difference
methods only surface elements have to be meshed. When the surface charges for
the individual voltage configurations are calculated the potential at any point
in space can be obtained by summing over the surface charges weighted by a
distance dependent $1/r$ scaling function.

\subsection{Single ion trapping}

\begin{figure}[htb]
  \begin{center}
\includegraphics[width= 1 \textwidth]{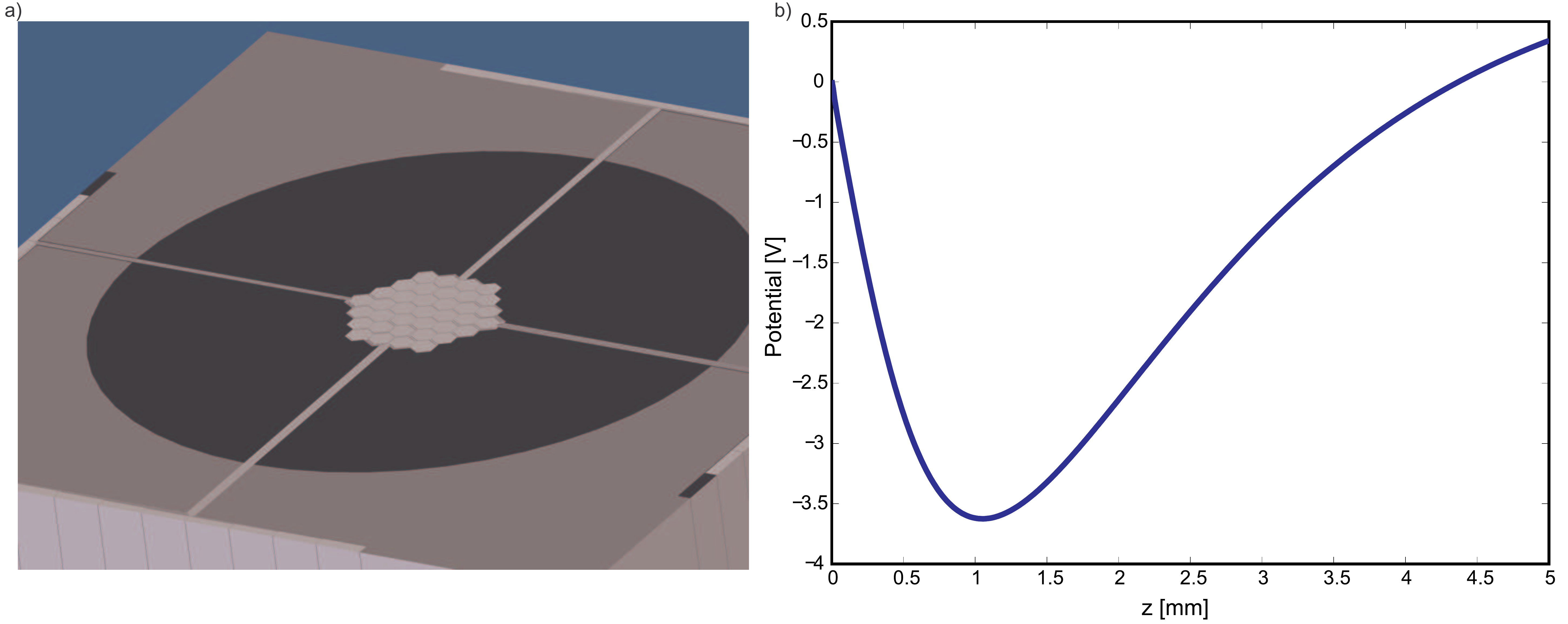}
\caption{a) Configuration with a large trapping volume and a deep confining
potential. All the hexagons are set to 0~V, the quartered ring around it serves as
the first ring electrode at -10~V and the rest serves as the outer ring electrode with
10~V. b) Axial trapping potential.} \label{bigtrap}
\end{center}
\end{figure}

For an initial trapping of ions, one usually chooses a large trapping volume.
To that end all control voltages of the pixels are set to 0~V while the outer
four circular segments are at -10~V and the four outermost electrodes at +10~V. The
resulting potential exhibits an axial frequency of 500~kHz and a depth of 3.5~eV,
ideally suited for trapping, cooling and observing a large cloud, see
Fig.~\ref{bigtrap}.

Starting from this initial setting, the inner electrodes are employed for single ion
trapping and we make use of the advanced configurations possible in a Pixel trap. Optimal potentials are obtained by applying regularization techniques \cite{SingerRMP2009}. If we supply the voltages $0$~V (white), $-0.2369$~V (black), $1.3171$~V (light grey), $-28.3125$~V (dark grey), $8.1845$~V (quartered ring, black), $10.2997$~V (outside plane, grey), see figures \ref{iontrapping}a) and \ref{bigtrap}a),  we would be able to reach a tight trap with an axial frequency of $\sim 1$~MHz and a trap depth of $\sim 0.7$~eV. The trap minimum is located at a distance of $\sim 0.2$~mm from the electrode surface, the anharmonicities are minimal and only lead to a broadening of the trap frequency well below a kHz for a particle as hot as $5$~K. We propose to load and laser-cool ions in the deep potential (see previous paragraph) and then alter the voltages to generate a harmonic well.

\subsubsection{Transporting a single ion} \label{transport sect}

\begin{figure}[htb]
\begin{center}
  \includegraphics[width=1\textwidth]{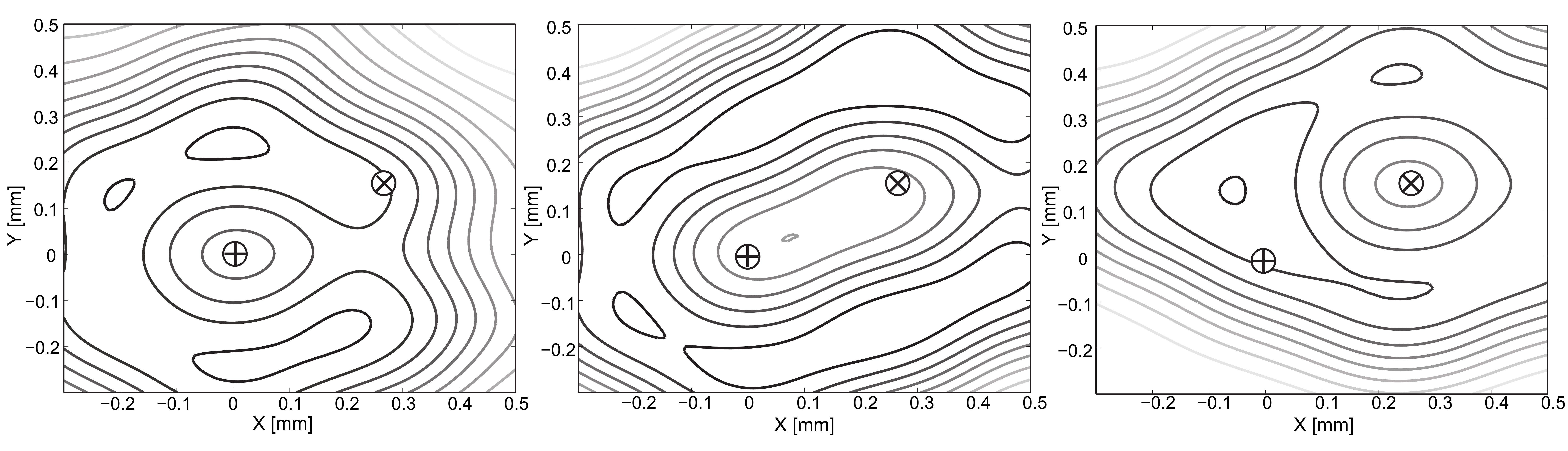}
  \caption{$\Phi_{el}(x,y)$ of the electrode configuration used at the start, in the middle and at the end of the transport, respectively, shown as a contour plots. The starting position of the transport is marked with $\bigoplus$, the endpoint with $\bigotimes$.\label{contour_transport}}
\end{center}
\end{figure}

Ions may be transported in the Pixel trap either (i) orthogonal to the plane
of electrodes, a transversal transport, or (ii) in a lateral transport parallel to
the trap electrode surface. In the case of (i) the pixels are arranged as rings on
equal potential, such that the position of the minimum of the axial potential varies
\cite{MAR09}. A possible application of a transversal transport would be the
determination of motional heating and decoherence rates as a function of
ion-surface distance, similar to attempts for dynamic Paul traps \cite{DES06}.
The lateral transport is sketched in Fig.~\ref{iontrapping} (a) to (c) (conceptually)
as well as Fig.~\ref{contour_transport} (radial potentials). Starting out with
a configuration that uses a single pixel as its center, we widen it up to an
elongated center, including adjacent pixels, then we tilt the potential in the
direction of the new center and finally finish with a single ion confinement
at the displaced position.  In our scheme, we intend to vary the electric
field slowly, such that the much faster cyclotron orbiting can follow
adiabatically. Alternatively, for other types of planar Penning traps, it has
been proposed \cite{CRI09} to apply a pulsed electric transversal field for a non-adiabatic transport. 

The ion is confined near the \emph{maximum} of the radial electric potential
$\Phi_{rad}(x,y)$ as the magnetron oscillation corresponds to an inverted
harmonic potential. Axialisation, an excitation of the ion's magnetron
frequency, centers it to $\Phi_{rad}^{max}$. Experimentally, the  outer four
segments, see Fig.~\ref{bigtrap} may be employed to generate a rotating wall
potential for this excitation \cite{POW02}.

The lateral transport of single ions may be readily expanded to a joining and splitting
operation for two ion crystals, a technique which has been so far limited to segmented
linear Paul traps \cite{ROW02,HUB08,BLA09}.

\subsection{Ring crystals and racetrack configuration}

In a linear Paul trap, typically ions arrange as linear ion strings, while two-
or three-dimensional arrangements of ions suffer from micro motion in the
dynamical trapping field since the micro motion vanishes only along one line,
where the RF field is zero \cite{HERS09,BERK98}. For crystal sizes on the
order of $10 \mu$m, this motion reaches amplitudes on the order of $1 \mu$m.
Here, a clear advantage of the purely static electric field of a Penning trap
becomes evident, because in our geometry two-dimensional ``artificial ion
crystals'' may be trapped by creating multiple trapping sites. The only
relevant motion left in this case is the cyclotron motion, which in the case
of an electron in a cryogenic trap (i.e. $T_e\sim4$~K, see \cite{BUSH08})
would have an orbit-radius of $r_{cyc}\sim15$~nm and in the case of a laser-cooled
Calcium ion (i.e. $T_{Ca}\sim1$~mK) a radius of $r_{cyc}\sim 65$~nm\footnote{An ion at room-temperature would of course show a quite large cyclotron orbit of $\sim35\mu$m}.

\begin{figure}[htb]
  \begin{center}
 \includegraphics[width=1\textwidth]{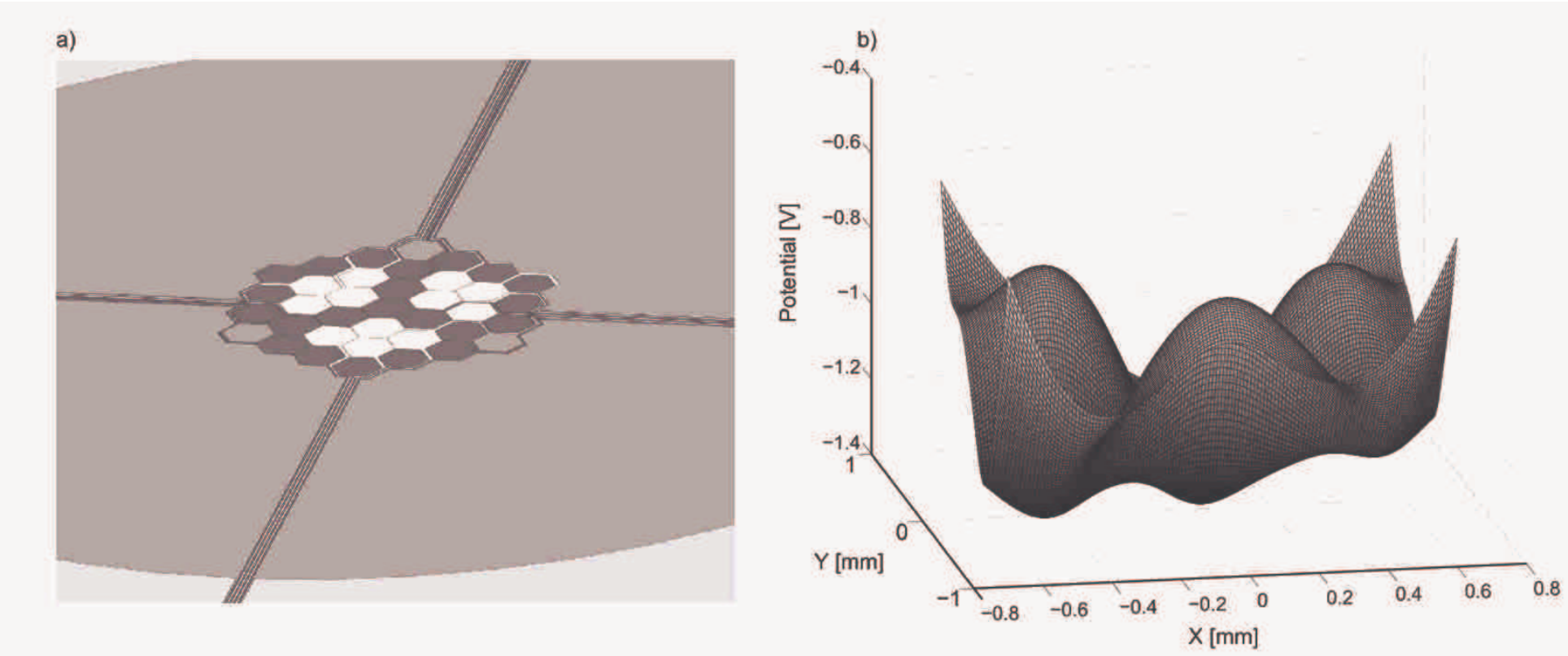}
\caption{a) Electrode configuration for an artificial ion crystal with three sites.
With voltages of 0.6~V: white, -4.0~V: dark grey, 0.5~V: light grey, we
reach a trap depth of 0.8~eV and axial frequencies near 800~kHz. b) xy-potential of the
trapping configuration shown in subfigure a), an artifical ion crystal with three ions.} \label{crystal}
\end{center}
\end{figure}

We highlight this with an arrangement of sites around a circle, trapping for
example three ions in predefined places. One example of such an electrode
configuration is shown in figure \ref{crystal}a.  The potential in the
xy-plane above such an electrode configuration is displayed in figure
\ref{crystal}b with three trapping sites. The interest of the Pixel Penning
trap is that various different kinds of such configurations may be converted
into each other by a time-dependent addressing of the pixel electrodes.

\begin{figure}[htb]
\begin{center}
\includegraphics[width=\linewidth]{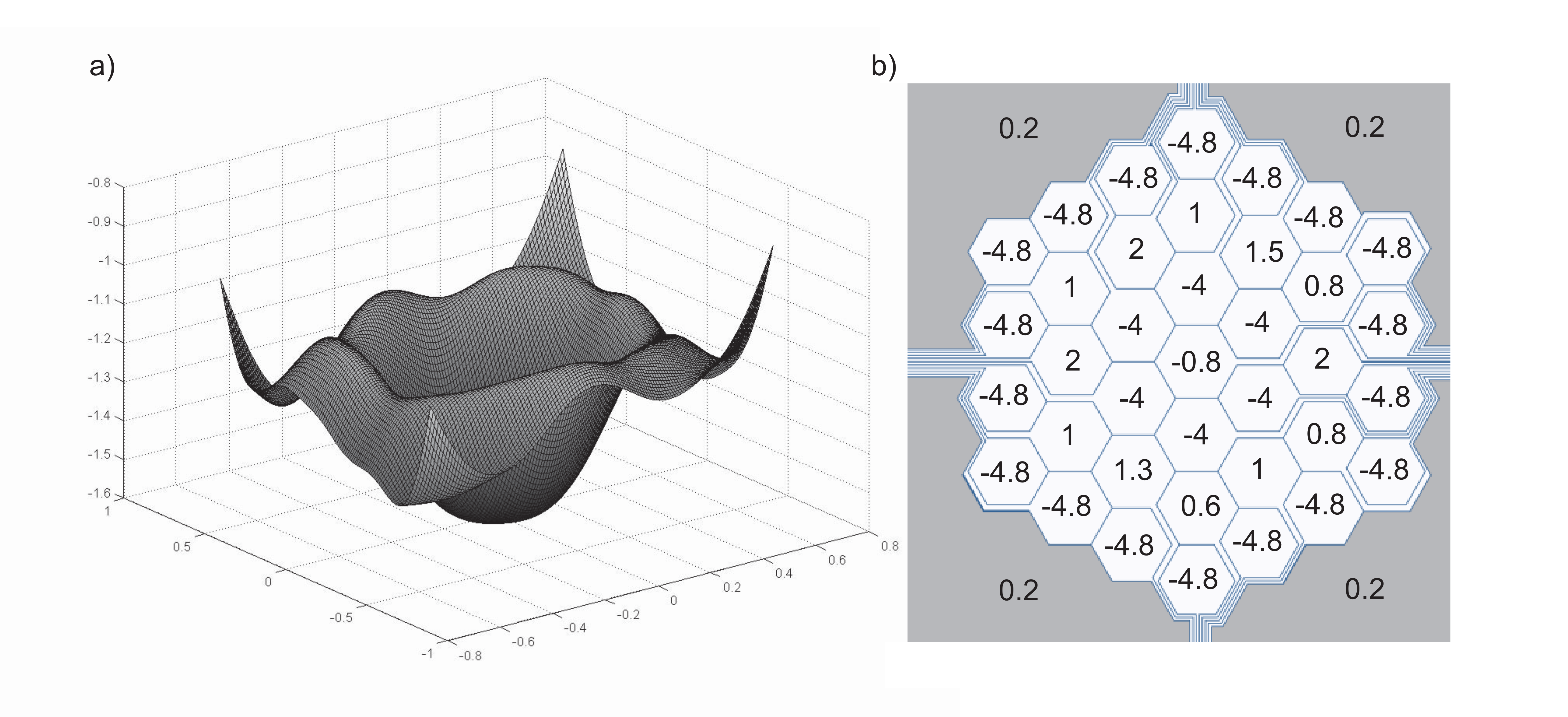}
\caption{a) Resulting potential $\Phi_{el}(x,y)$ for a racetrack configuration. b) Map of the voltages needed to create this potential (unit is Volts). The differences in the individual voltages on the actual racetrack are due to slightly varying sizes of the pixels.}
\label{racetrack}
\end{center}
\end{figure}

The colloquial term ``racetrack'' is commonly used for a configuration in which the
trapping zone is in the form of a ring. Radio frequency ion traps have been used to
hold linear crystals in a racetrack configuration \cite{BIR92}, inspired by confined
charged particles in accelerator rings. Magnetic traps with additional radio-frequency
fields have been proposed for racetrack configurations for interferometric purposes
\cite{LES07}. Optical potentials for trapping neutral atoms \cite{KAP01,RYU07} are
yet another option. Our Penning trap approach does not require any time dependent
fields, advantageous with regard to heating and decoherence, and technically less
demanding. In the Pixel Penning trap we may realize a 580~$\mu$m diameter racetrack, see
Fig.~\ref{racetrack}.

\section{Fabrication of the Pixel trap}

\begin{figure}
    \begin{center}
    \includegraphics[width=0.7 \linewidth]{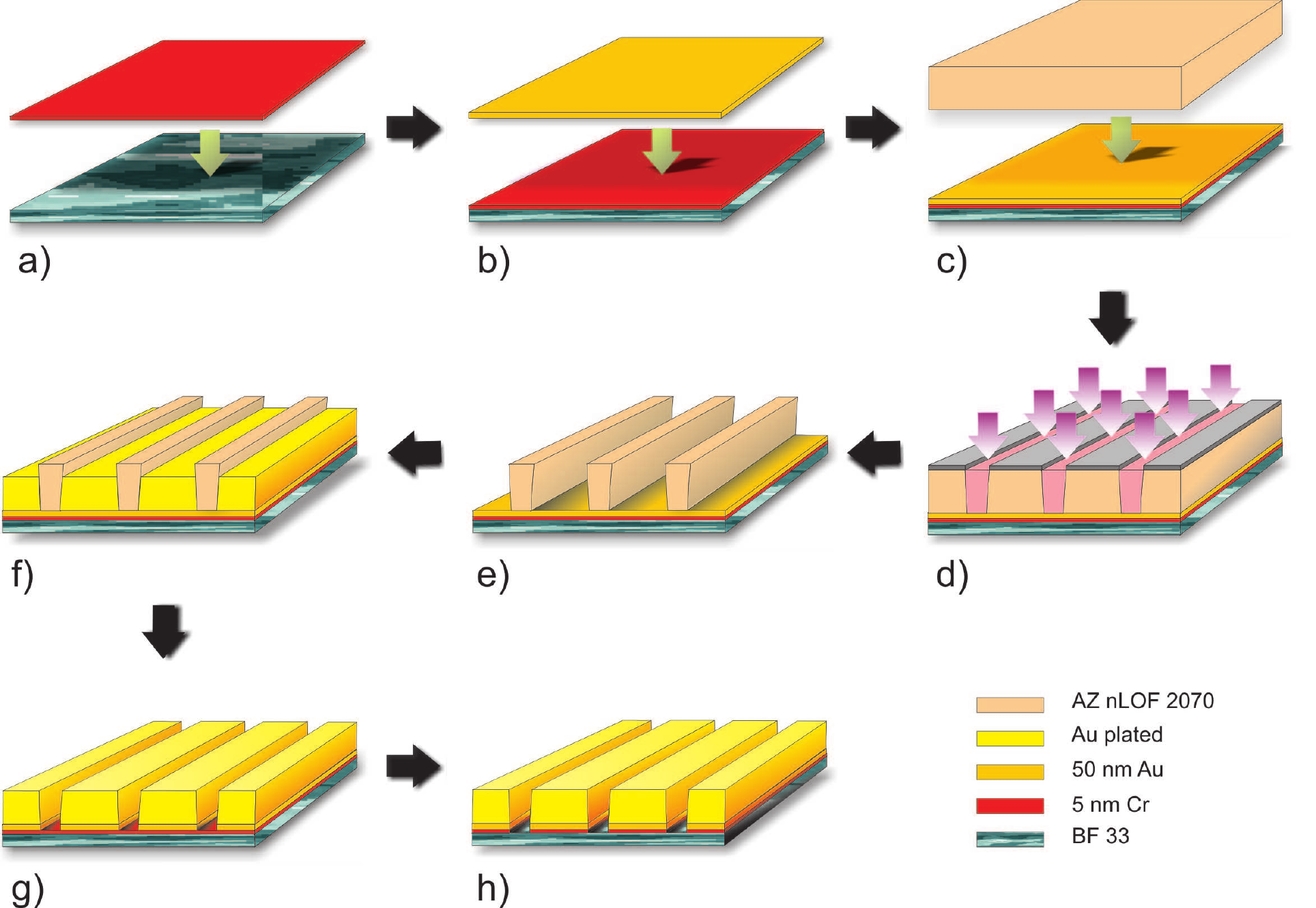}
    \caption[Fabrication scheme]{Simplified scheme of the fabrication process.
    a) Evaporation of the adhesion Cr layer on BF$^{\circledR}\,33$ wafer,
    b) Evaporation of the starting Au layer on Cr/BF$^{\circledR}\,33$, c) Spin coating
    of AZ nLOF 2070 resist, d) Photolitography: exposure to UV light,
    e) Development, f) Electroplating of Au, g) Stripping of resist and wet-etching of Au seed layer,
    h) Wet-etching of Cr.}
    \label{Fabscheme}
    \end{center}
\end{figure}

\begin{figure}
    \begin{center}
    \includegraphics[width=0.6\textwidth]{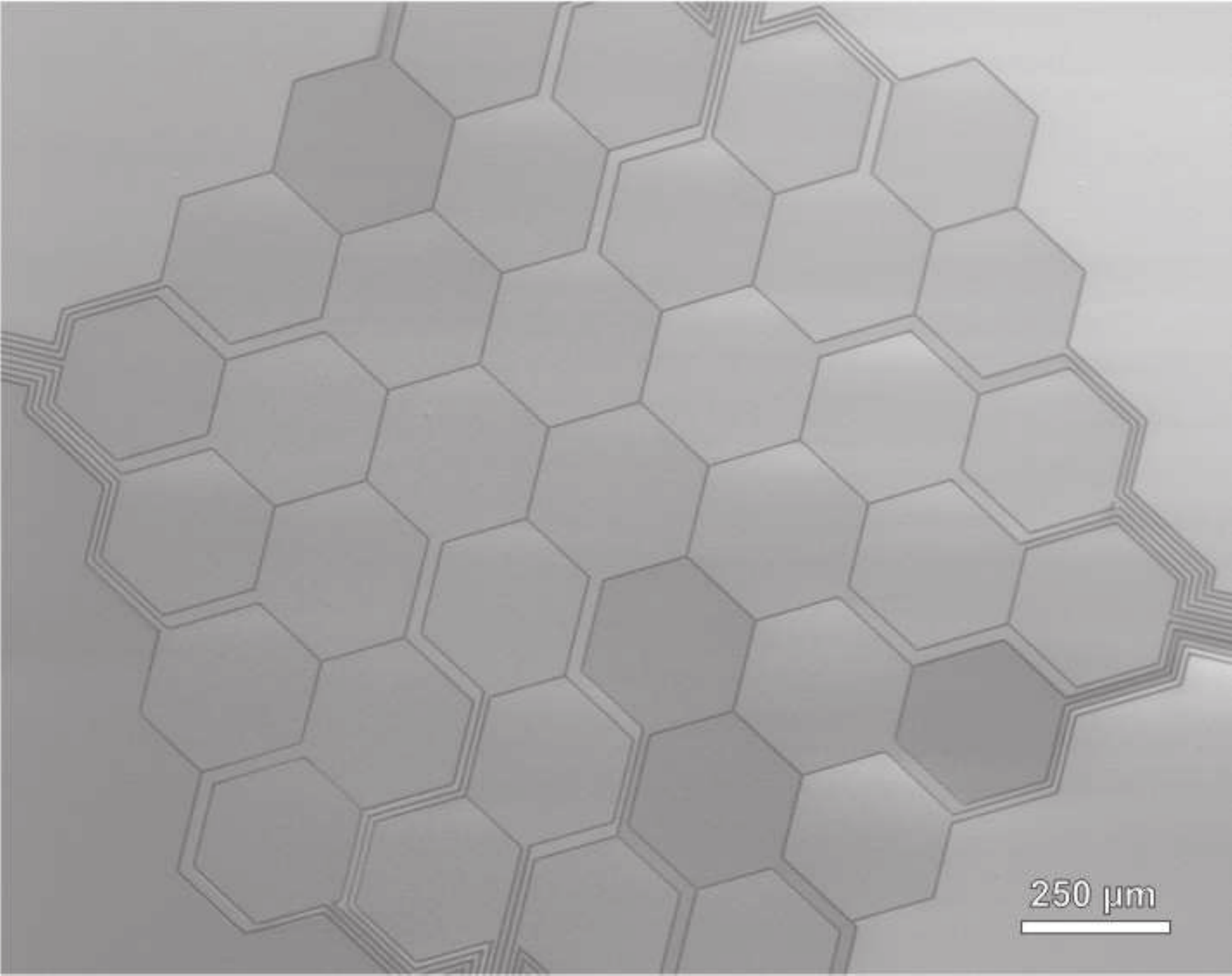}
    \caption[scanning electron microscopy]{SEM micrograph of the
    Gold-on-sapphire Pixel Penning trap}.
    \label{SEM}
    \end{center}
\end{figure}

Starting with the CAD file of the trap design, we realize the fabrication of
planar Penning traps by three-step clean room processing: photolithography,
electroplating and etching. The fabrication is schematically presented in
Fig.~\ref{Fabscheme}. A polished borosilicate glass wafer serves as substrate.
We have achieved similar results with 2" sapphire wafers of thickness
$0.50$~mm, which has the advantage of better heat conduction, even at cryogenic
temperatures. In order to ensure proper adhesion of the Au electrodes to the
substrate an adhesion promoter layer is mandatory. We use a 5~nm Cr layer
deposited on the wafer via thermal evaporation. Furthermore, to enhance Au
electrodeposition a 50~nm Au seed layer is also thermally deposited on top and
a negative resist \footnote{AZ${}^{\circledR}$~nLOF~2070} is added.
Photolithography is then carried out. The mask resist lines which define the
gaps between electrodes show a width of 4.0~$\mu$m. Electrodes are grown up to
4.0~$\mu$m in a Au plating bath \footnote{mixture of Enthone GRC complex and
Enthone microfab Au-100B}. This means we achieve an aspect ratio of 1 between
depth and width of the gaps between the electrodes, shielding inter-electrode
insulating substrate gaps from the ion position. After electroplating the
resist mask is stripped off. The Au seed layer is removed (using KI/I$_2$ gold
etchant) and a final Cr etching step is performed. Fig.~\ref{SEM} shows an
Scanning Electron Microscopy micrograph of the inner center of the Pixel trap.

We dice the wafers and ball bond the chip electrodes to an 84-pin Kyocera chip carrier
using 25~$\mu$m Au wires. Under UHV conditions with $10^{-7}\,\mathrm{mbar}$ we
investigate the limit of the applied electric field by raising slowly the
applied voltage. We observe a breakdown at approximately 180~V for 1.2~$\mu$m
to 720~V for 4.0~$\mu$m gap width, respectively, far above the required trap voltages.

\section{Outlook}
As a next step, we plan experimental tests with single trapped $^{40}$Ca$^+$ ions in the
Pixel trap in order to particularly optimize the control of the transport processes
 \cite{KIRK,CHIAR08}. The Pixel trap opens a way to miniaturized Penning traps,
and once the coherence properties have been tested with single ions close to
surfaces, one might further reduce the spatial dimensions of the surface
electrodes to about 10~$\mu$m, including a dual layer technique
\cite{AMI08}. In a planar arrangement of trapped atoms one may perform
entanglement gate operations which are mediated by an inhomogenious magnetic
field \cite{JOH09,WUN09}. The interaction Hamilton operator
H$_{spin}=\hbar/2\sum_{n<m} J_{nm} \sigma_{z,n}  \sigma_{z,m}$ denotes the
coupling of two spins, where the strength of the coupling $J_{nm}$ can be
controlled by the distance of the atoms and the magnetic gradient field. With
ion-ion distances of a few 10$\mu$m, and realistic magnetic gradient
fields of $10$ to $50$~T/m one will reach a coupling strength J$_{nm}$ of kHz for
the mutual effective spin-spin interaction. We aim for scaling up the number of
qubits which participate in the generation of cluster states (see
Ref.~\cite{WUN09} for details) using a two-dimensional instead of a linear ion
crystal. Further applications in quantum science would be experimental tests
of quantum state transfer protocols \cite{CHR2004,BOS2007}, which have been
proposed but not realized experimentally so far. The two-dimensional ion
crystals would also facilitate quantum simulation because the spin-spin
distance and the geometry could be controlled by the trapping fields even more
flexibly than in the situation of ultra-cold atoms in optical lattices
\cite{BLO2008}.


\section*{Acknowledgements}
A. B.-S. thanks CONACYT for financial support through the scholarship ID~206267. We acknowledge financial support by the German science foundation DFG within the SFB/TRR-21, the European commission within MICROTRAP (Contract No. 517675) and the excellence program of the Landesstiftung Baden-W\"urttemberg.

\end{document}